\documentclass[information,article,accept,pdftex,moreauthors]{Definitions/mdpi} 
\firstpage{1} 
\pubvolume{1}
\issuenum{1}
\articlenumber{0}
\pubyear{2025}
\copyrightyear{2025}
\datereceived{ } 
\daterevised{ } 
\dateaccepted{ } 
\datepublished{ } 
\hreflink{https://doi.org/} 

\usepackage{rotating}

\Title{Leveraging Machine Learning Techniques to Investigate Media and Information Literacy Competence in Tackling Disinformation}

\TitleCitation{Title}

\Author{José Manuel Alcalde-Llergo $^{1,4}$\orcidA{}, Mariana Buenestado Fernández $^{2}$\orcidB{}, Carlos Enrique George-Reyes $^{3}$\orcidC{}, Andrea Zingoni $^{4}$\orcidD{}, and Enrique Yeguas-Bolívar $^{5}$*\orcidE{}}

\AuthorNames{Firstname Lastname, Firstname Lastname and Firstname Lastname}

\isAPAStyle{%
       \AuthorCitation{Lastname, F., Lastname, F., \& Lastname, F.}
         }{%
        \isChicagoStyle{%
        \AuthorCitation{Lastname, Firstname, Firstname Lastname, and Firstname Lastname.}
        }{
        \AuthorCitation{Lastname, F.; Lastname, F.; Lastname, F.}
        }
}

\address{%
$^{1}$ \quad Dept. of Computer Science, University of Córdoba, Córdoba, Spain; jmalcalde@uco.es\\
$^{2}$ \quad Dept. of Education, University
of Córdoba, Córdoba, Spain; marianabf@uco.es\\
$^{3}$ \quad Universidad Politécnica Metropolitana de Hidalgo, México; cgeorge@upmh.edu.mx\\
$^{4}$ \quad Dept. of
Economics, Engineering, Society and Business Organization, University of
Tuscia, Viterbo, Italy; andrea.zingoni@unitus.it\\
$^{1}$ \quad Dept. of Computer Science, University of Córdoba, Córdoba, Spain; eyeguas@uco.es}

\corres{Correspondence: eyeguas@uco.es}

\usepackage{comment}

\usepackage[normalem]{ulem} 

\abstract{This study develops machine learning models to assess Media and Information Literacy (MIL) skills specifically in the context of disinformation among students, particularly future educators and communicators. While the digital revolution has expanded access to information, it has also amplified the spread of false and misleading content, making MIL essential for fostering critical thinking and responsible media engagement. Despite its relevance, predictive modeling of MIL in relation to disinformation remains underexplored. To address this gap, a quantitative study was conducted with 723 students in education and communication programs using a validated survey. Classification and regression algorithms were applied to predict MIL competencies and identify key influencing factors. Results show that complex models outperform simpler approaches, with variables such as academic year and prior training significantly improving prediction accuracy. These findings can inform the design of targeted educational interventions and personalized strategies to enhance students’ ability to critically navigate and respond to disinformation in digital environments.
}

\keyword{Media and information literacy; disinformation; higher education; machine learning} 

\begin{document}


\section{Introduction}
\label{sec:intro}

The current international political agenda underscores the need to develop and implement policies, action plans, and strategies aimed at promoting MIL. These efforts seek to raise public awareness, strengthen prevention capacities, and foster resilience against disinformation and the spread of misinformation~\citep{MarcosVilchez24}. The relevance of these issues is reflected in the UN General Assembly Resolution (2021), which calls on Member States and stakeholders to design and implement specific measures to promote MIL. Moreover, it highlights the importance of increasing awareness, enhancing prevention capacity, and reinforcing resilience in the face of disinformation and false information. This approach is also recognized as a key solution in UNESCO's Windhoek +30 Declaration (2021), where one of the three core principles to counter disinformation is fostering critical thinking among citizens, promoted through MIL initiatives.

However, a report from the same organization warns that young people, due to their constant interaction with digital platforms, are particularly vulnerable to disinformation. Approximately 70\% of young people globally are connected online but still lack solid and sustainable training in MIL to tackle the challenges of the digital age~\citep{UNESCO23}. The importance of adapting the guidelines of international organizations as a foundation for innovating the development of practices that strengthen professional training and address specific issues related to education and communication fields is often emphasized~\citep{MateusDeOro24}.

In the development of good MIL practices to counter disinformation, the scientific literature has sought to identify a competency profile that, through processes of social and professional transfer, allows for the selection of training proposals that positively impact the teaching culture and student body~\citep{RomeroRodriguez19, AlcoleaDiaz20}.

In recent years, MIL has evolved beyond its traditional focus on media access, critical consumption, and message production to address new challenges posed by digital misinformation and disinformation~\citep{UNESCO23}. While MIL generally promotes critical thinking and informed engagement with media, its application in the context of disinformation is distinctive because it requires not only the evaluation of information credibility but also an understanding of the intentional manipulation of content and its socio-political consequences. Disinformation represents a deliberate and systematic distortion of information, designed to mislead or polarize audiences, which demands higher-order cognitive, ethical, and civic competencies. Therefore, exploring MIL specifically as a means to counter disinformation highlights a more proactive and defensive dimension of literacy, one that equips individuals not just to interpret media critically, but to actively resist and respond to manipulative information flows in digital environments.

MIL against disinformation, like other topics in educational sciences, can be addressed through more advanced research designs and analytical techniques. These approaches open numerous opportunities in the educational field, in line with the strategic recommendations established in the Beijing Consensus on Education~\citep{UNESCO19}. Such advancements have the potential to transform educational processes through improved monitoring, evaluation, and research~\citep{OECD21}. In particular, Machine Learning (ML) offers a wide range of algorithms capable of analyzing and interpreting complex data with great precision, facilitating the creation of predictive models that adapt to individual circumstances and the specific progress of each participant in the study~\citep{Zhai21}.

While recent advances in generative artificial intelligence have expanded the possibilities for content creation and educational innovation, they also raise concerns about reliability, bias, and hallucinated outputs that can distort information and amplify misinformation. These challenges underscore the need for stronger MIL skills to critically assess and responsibly engage with artificial intelligence-generated content. In contrast, the ML approach adopted in this study serves a different purpose: it functions as an analytical framework for identifying patterns and predicting competencies based on empirical data, without generating or altering information. By leveraging ML in this way, the study contributes to evidence-based educational research while remaining aligned with the ethical imperatives of fostering critical, data-informed literacy among students.

Various ML techniques have been employed to address the challenge of detecting disinformation on social networks, a topic that has received considerable attention, as evidenced by several literature reviews~\citep{Aimeur23, Sanaullah22, Korkmaz19}. However, the application of these techniques remains in an early stage as methodological proposals in the field of research on MIL to combat disinformation~\citep{FastrezLandry23}. In this context, the present article provides a valuable and original contribution to the scientific literature by analyzing the performance of ML models in predicting sociodemographic variables of future professionals in the fields of education and communication, specifically focusing on their perception of MIL against disinformation. By exploring the intersection between ML and the prediction of sociodemographic variables, this study expands our understanding of the potential of AI techniques to reveal relevant information and support informed decision-making in the educational field.

The article is organized as follows: Section~\ref{sec:intro} presents the object of study and its relevance. Section~\ref{sec:background} reviews the background and analyzes previous studies related to MIL against disinformation, as well as the associated variables and the use of ML in the educational field. Section~\ref{sec:methodology} details the methodology, including the research design, study variables, data collection instrument, and participant characteristics. Section~\ref{sec:experimentation} outlines the experimental design, evaluating the performance of ML models in three key tasks: classification of the knowledge branch, selection of the most relevant variables, and regression of key factors. Section~\ref{sec:results} presents the results obtained during the study. Section~\ref{sec:discussion} provides a comprehensive discussion of the findings, highlights their implications, and outlines the study’s limitations. Finally, Section~\ref{sec:conclusions} concludes the article by highlighting the main findings and proposing directions for future research.

\section{Background, Variables and Research Model}
\label{sec:background}

The spread of disinformation in digital environments has highlighted the importance of MIL as a key competency for thoughtfully engaging with media content. Research suggests that MIL is not only about accessing information but also about developing ethical and analytical thinking skills to analyze, evaluate, and create content responsibly. This section reviews previous studies on MIL and its relationship with disinformation, focusing on three key aspects.

First, it explores how MIL empowers individuals to identify misinformation, detect biases, and navigate the complexities of the modern information landscape. Then, the demographic and academic factors, such as gender, age, academic background, and previous training, that influence MIL levels are examined. Finally, it discusses how ML methods have been used to assess and predict MIL competencies, offering insights into their role in educational contexts.

\subsection{MIL against Disinformation}

MIL is a set of key knowledge, skills, and attitudes that enable individuals to access, analyze, evaluate, and create media and informational content critically and responsibly. Scientific studies on this topic indicate that MIL is not only about technical competencies to access information but also involves a critical and ethical capacity to understand the messages conveyed by the media, as well as the contexts in which they are produced and distributed~\citep{Knaus20}.

It has been shown that people with strong MIL are capable of discerning between reliable and unreliable sources, identifying biases in content, and understanding the intentions and objectives of information producers~\citep{Gross25, Huang24}. This ability is especially crucial in a world where disinformation, fake news, and algorithms prioritizing sensationalist and polarizing content dominate many platforms~\citep{Cooper19}.

Experts emphasize that MIL empowers individuals, granting them control over the information they consume and produce~\citep{Aguaded15, Hobbs22}. Instead of being passive consumers, people should be active participants who evaluate, question, and create content that promotes a more democratic, transparent, and respectful society.

The MIL approach requires disciplinary convergence and employs a conceptual framework grounded in various fields of knowledge, drawing inspiration particularly from communication studies, sociology, psychology, political science, engineering, and education sciences~\citep{Landry20}. Specifically, educommunication is based on the interconnection between two traditionally separate intellectual fields, education and communication, which now converge to address a new social and educational need~\citep{Aguaded19}.

\cite{Osuna18} highlights the need for communicators to take on greater social responsibility and for educators to integrate the relevance of media into their pedagogical methods.

\subsection{Associated Variables}

A wide range of variables has been examined in the literature to explain differences in MIL levels, particularly in the context of disinformation. Representative factors commonly reported include gender, age, academic year, academic field (education or communication), and prior training in media literacy or disinformation. In addition, MIL has been associated with broader contextual and behavioral dimensions, such as learners’ socio-cultural background, digital engagement, and patterns of media consumption, which together shape how individuals acquire and apply critical information skills. These variables were selected because they have been repeatedly associated with variations in critical evaluation, digital engagement, and exposure to media education, allowing the analysis to focus on demographic and academic factors most consistently linked to MIL in previous studies.

Regarding gender, although women tend to be more reflective than men, they are still less likely than men to frequently verify the accuracy of information in the media \citep{Golob21}. Additionally, female university students have greater difficulty identifying fake news when it comes from media sources with a humorous tone \citep{HerreroDiz19}. However, other studies report no significant gender differences or even higher verification rates among women \citep{Almenar21}. Furthermore, studies indicate that women are more likely to engage with social media content in ways that can increase exposure to misinformation, highlighting a potential gender-related susceptibility to false information \citep{Mansoori23}. These findings underscore the complexity of gender-related patterns in MIL and the need for further research to clarify these relationships.

The studies by~\cite{Bouleimen23} and~\cite{Pennycook20} hypothesized that online searches are more likely to lead young people to validate true news than to refute false ones. However, they concluded that age does not influence their ability to identify false information. Similarly,~\cite{RomeroRodriguez19} reported that levels of media competence are not linked to specific age groups. While associating university students with the terms ``digital natives'' and ``digital immigrants'' may be useful to describe instrumental access to platforms and digital competencies, these terms do not guarantee an adequate level of media literacy or critical media consumption~\citep{Avello21}. 

The study by~\cite{HerreroCuriel22} also highlights differences between first-year and final-year students in communication-related degrees concerning the effectiveness of tools acquired during their studies, emphasizing the need to review and improve university training strategies. 

On the other hand,~\cite{HerreroDiz19} emphasized significant differences in media competence in the face of disinformation between university students in communication and education. In particular, education students face greater difficulties in correctly interpreting aspects such as the accuracy, authorship, and purpose of texts. 

Lastly,~\cite{Cherner19} discussed that adequate training in media literacy is essential for individuals to develop the competencies necessary to critically evaluate and verify information. People with training in this area face fewer difficulties in interpreting content, a conclusion supported by studies such as those by~\cite{Kahne17} and~\cite{Fajardo15}, which emphasize from different perspectives that media literacy is the only effective tool to protect society against disinformation. However,~\cite{Adjin22} warns that one-off training, while positive, is insufficient to consolidate strong critical skills. In this sense, a continuous and consistent educational approach is essential to enable students to identify and evaluate information more effectively. 

\subsection{Application of ML Techniques in Educational Perceptions}

ML is a subfield of AI that focuses on developing models and algorithms capable of learning and making predictions or decisions based on data, without requiring explicitly programmed instructions for each task~\citep{Chen20}. The literature review by~\cite{Hilbert21} on the use of ML in educational sciences concludes that, although the impact of these methods in research and their practical applications in this field is still limited, it is constantly growing and is significantly transforming traditional approaches to educational research. Scientific evidence has shown that one of the main advantages of ML over traditional approaches, such as linear regression, is its ability to incorporate non-parametric components, making it more adaptable and versatile in practical applications as a predictive model~\citep{Ersozlu24,Gibson17}.

In educational contexts, this adaptability allows ML algorithms to analyze large amounts of heterogeneous student data to detect individual learning trajectories and predict areas where students may struggle. By identifying these patterns, ML systems can recommend or even generate content, assessments, and feedback that match each learner’s profile. In this way, ML facilitates the design of personalized curricula that adapt dynamically to students’ needs, optimizing teaching strategies and fostering more effective and equitable learning progress~\citep{Munir22,Luan21}.

ML has found various applications in the educational field, including the exploration of more subjective aspects, such as students' perceptions of their level of educational competencies. Examples of this include competency in computational thinking~\citep{Tan23}, competency in open education~\citep{Ibarra-Vazquez24,Estrada-Molina24}, digital teaching competency~\citep{Forero-Corba24}, competency in complex thinking~\citep{Ibarra-Vazquez24}, and competency in global citizenship~\citep{Bernardo22,Miao23}.

Some relevant studies are related to the topic under study, although the application of ML techniques in research on MIL is still in a relatively nascent phase. The study by~\cite{Reddy19} focuses on predicting the level of media literacy through the use of classification algorithms, such as random forest (RF), decision trees (DT), and support vector machines (SVM), with the aim of building predictive models. The results show that media literacy skills among students are limited, with 61.4\% of participants at low literacy levels. Two key factors for improving these skills are the interpretation of media messages and the use of various sources and devices.

On the other hand, the study by~\cite{Wusylko24} analyzes student participation in media literacy campaigns on TikTok using ML techniques. The study explored how topic modeling, sentiment analysis, and network analysis can provide valuable insights into students' interactions with these campaigns. In a different approach,~\cite{MohdNadzir23} proposed a predictive model of digital media literacy in the context of distance education. During the data analysis phase, they selected five ML algorithms as classifier models: logistic regression (as a baseline model), k-nearest neighbor (KNN), RF, SVM, and multilayer perceptron. Additionally, they employed five cross-validation configurations to evaluate the models. The comparison of results was conducted using metrics such as accuracy, precision, recall, and F-measure. The best model, based on the SVM algorithm, demonstrated the highest accuracy: 82.9\% for model learning based on generation and 41.4\% for device-based learning.
This study identified interesting patterns according to the generation of students, revealing that different generations employ digital literacy in distinct ways. Thus, the proposed predictive model demonstrated that the generation of students is a useful indicator for evaluating their level of digital media literacy, especially in the case of distance education students.

\section{Methodology}
\label{sec:methodology}

This section presents the methodology used in this research, including the research design, study variables, the instrument employed for data collection, and the participants involved. The focus is on understanding how MIL impacts the ability to critically engage with disinformation in educational contexts, particularly for future educators and communicators. The research design and corresponding data collection methods
were therefore structured to develop and validate predictive models of MIL competencies based on sociodemographic and academic factors, allowing for the identification of variables that most strongly influence students’ capacity to address disinformation.

\subsection{Research Design and Study Variables}

The study followed a quantitative design and employed a survey-based approach to assess students’ MIL in the context of disinformation. From a temporal perspective, this is a cross-sectional study, as data were collected at a single point in time.
The selection of study variables is directly informed by the literature reviewed in Section~\ref{sec:background}. Prior research has shown that MIL competencies are influenced by demographic and academic factors, such as gender, age, academic year, field of study, and previous training in media literacy or misinformation. These factors were therefore included as independent (predictor) variables.
The dependent (criterion) variables correspond to the core dimensions of MIL identified in previous studies and discussed in Section~\ref{sec:background}, namely: (1) knowledge about misinformation, (2) skills and behaviors in responding to misinformation, and (3) attitudes of commitment and responsibility toward misinformation. This conceptualization aligns with empirical findings showing that effective MIL requires not only cognitive understanding but also behavioral competencies and ethical attitudes. By linking predictor and criterion variables to established literature, the study ensures that the analyses are grounded in the theoretical and empirical context of MIL in the face of disinformation.

\subsection{Instrument}\label{sec:instrument}

For this research, a questionnaire titled \textit{Perception of Future Edu-Communicators on Disinformation} was designed. The instrument consists of 25 items organized on a 4-point Likert scale (1 = None, 2 = Little, 3 = Quite a bit, 4 = A lot). Its structure is divided into three main sections:
\begin{itemize}
    \item \textbf{Sociodemographic variables:} Five items collect information on gender, age, academic year, academic field, and prior training in MIL and disinformation.
    \item \textbf{Theoretical dimensions of MIL in relation to disinformation:} Five items for each theoretical dimension, covering knowledge, skills, and attitudes toward disinformation.
    \item \textbf{Responsibility toward disinformation:} Five items assess the perception of responsibility as future edu-communicators in combating disinformation.
\end{itemize}

Regarding the origin of the instrument, no existing questionnaire in the literature addressed perceptions of MIL in relation to disinformation for both future educators and communicators while aligning with our research objectives and hypotheses. Therefore, an \textit{ad hoc} questionnaire was developed based on a review of the literature and documents from prestigious organizations. Concerning the structure of the instrument, the three dimensions (knowledge, skills, and attitudes) are intrinsic to the concept of competence~\citep{OECD05}. Moreover,~\cite{UNESCO23} emphasizes that digital communication poses challenges such as disinformation. To address these challenges and foster a media- and information-literate citizenship, it is essential to promote media education and critical thinking, particularly during the initial training of education and communication professionals. Taking this critical observation into account, the dimension of responsibility was added. The items corresponding to the different dimensions were identified from previously validated questionnaires used in studies such as those by~\cite{CatalinaGarcia19},~\cite{Figueira19}, and~\cite{HerreroCuriel22}. Subsequently, these items were reformulated and adapted to fit the specific characteristics of the participants in this study.

To assess content validity, the expert judgment method was employed. A purposive sample of five specialists was selected using the convenience sampling method. These experts met the following criteria: doctoral studies in education or communication, experience in research projects, scientific output related to media literacy and disinformation, and at least twelve years of research experience. Regarding Kendall's W test, similar results were obtained. For the clarity criterion, a value of $p < .000$ was observed; for the coherence criterion, $p < .032$; and for the relevance criterion, $p < .003$, reflecting statistical significance in all cases. Quantitative evaluations did not provide relevant information for improving the instrument. However, adaptations and modifications were made to the wording of some items based on qualitative evaluations or comments from participants in the expert judgment process.

\subsection{Participants and Data Collection}\label{sec:participants}
The sample consisted of a total of 723 university students from two fields of study: 435 from Education degrees and 288 from Communication degrees. The study was conducted in a region in northern Spain, where both degrees share the Media and Information Literacy competency in their curricula. Regarding gender, 201 participants were men and 522 were women. Among the sociodemographic data, a significant tendency was observed for women to specialize in fields related to communication and education, a phenomenon reflected in the participants' profile~\citep{CatalinaGarcia19, MarinSuelves21}.

The age distribution of the participants was as follows: 277 students were between 17 and 19 years old, 384 were between 20 and 22 years old, and 62 were over 23 years old. Additionally, participants came from all four academic years, distributed as follows: 209 students were in their first year, 161 in their second year, 161 in their third year, and 192 in their fourth year. Regarding prior training in disinformation, 218 students reported having received training on this topic, while 505 indicated they had not.

The inclusion criteria for the sample were enrollment in one of the two aforementioned fields of study and access to devices with an internet connection to respond to the online questionnaire. A convenience sampling method was used, as participants were selected from students enrolled in courses taught by the researchers.

The participants have been informed about the purpose of the study, and their responses are anonymous and confidential. All have explicitly given their consent to participate in this research voluntarily.

\subsection{Procedure and Data Preprocessing}
The data collected were processed using SPSS v.25 and Python. In the first phase, an exploratory factor analysis was conducted to assess the validity and reliability of the attitude questionnaire, aiming to identify the empirical structure of the data. In the second phase, the study's objectives were addressed. 

Before conducting the experiments, a data preprocessing step was performed. This process included normalization and encoding of categorical variables using the One-Hot Encoding technique. Additionally, the final scores for each subject were computed for the four categories: knowledge, skills, attitudes, and responsibilities. As a result, the final dataset consists of 30 features.

\section{Experimentation}
\label{sec:experimentation}
This section describes the experimental design of this study. Specifically, the objective is to analyze the performance of different ML models in three key tasks:

\begin{itemize}
    \item \textbf{Classification of the Knowledge Branch:}  
    This task evaluates whether students’ MIL profiles differ systematically between the two academic areas considered: Education and Communication.  
    The purpose is not merely to label students but to identify discipline-specific patterns that may influence the development of MIL competencies.  
    This helps reveal latent relationships in the dataset that are not easily observable through descriptive statistics alone.
    
    \item \textbf{Selection of the Most Relevant Variables:}  
    Different feature selection techniques were applied to determine which variables contribute most to classification performance and to interpret their educational significance in predicting MIL outcomes.

    \item \textbf{Regression of the Key Factors:}  
    Regression models were trained to estimate the scores students would achieve in each MIL dimension (Knowledge, Skills, Attitudes, and Responsibility).  
    This allows the identification of which sociodemographic and academic factors most strongly predict each competence and the extent of their influence.
\end{itemize}

All algorithms and configurations used for the different tasks were implemented locally with open-source Python libraries (\texttt{scikit-learn}~\citep{scikit-learn} and \texttt{LightGBM}~\citep{lightgbm}).

The experiments were conducted using the dataset derived from the questionnaire and data collection described in Section~\ref{sec:instrument} and Section~\ref{sec:participants}. After preprocessing, each row in the dataset represented one student and included both sociodemographic information and aggregated scores for the four MIL dimensions: Knowledge, Skills, Attitudes, and Responsibility. In total, 723 valid responses were available, considering 30 numerical features for input into the different models.
The following subsections provide a detailed breakdown of each stage in the experimental process.

\subsection{Classification of the Knowledge Branch}\label{sub:class}

To conduct the experimentation for the prediction of the knowledge branch, various ML algorithms and different hyperparameter configurations have been considered. The algorithms and configurations are as follows:

\begin{itemize}
    \item \textbf{Support Vector Machine (SVM):} Different configurations were explored by varying the regularization parameter \( C \) among \{0.1, 1, 10, 100\}, the kernel function among \{\textit{radial basis function (RBF)}, \textit{linear}, \textit{polynomial}\}, and the parameter \(\gamma\), which controls the influence of training samples in defining the decision boundary. Instead of manually selecting \(\gamma\), two widely used and theoretically justified heuristics were employed:
    \begin{itemize}
        \item \(\gamma = \frac{1}{n_{\text{features}} \cdot \text{Var}(X)}\), where \( n_{\text{features}} \) is the number of input features and \( \text{Var}(X) \) is their variance.
        \item \(\gamma = \frac{1}{n_{\text{features}}}\), which only depends on the number of features.
    \end{itemize}
    
    \item \textbf{Decision Tree (DT):} Experiments were conducted by varying the maximum tree depth among \{\textit{unrestricted}, 5, 10, 20\} and the minimum number of samples required to split a node among \{2, 5, 10, 20\}.
    
    \item \textbf{Random Forest (RF):} Different numbers of DTs (estimators) were considered, choosing among \{10, 50, 100\}. Additionally, maximum tree depths \{\textit{unrestricted}, 5, 10, 20\} and minimum sample sizes for node splitting \{2, 5, 10, 20\} were explored.
    
    \item \textbf{Light Gradient Boosting Machine (LightGBM):} The model was evaluated with different numbers of boosting iterations (estimators) among \{50, 100, 200\}, maximum depths among \{\textit{unrestricted}, 5, 10\}, and learning rates among \{0.01, 0.1, 0.2\}.
    
    \item \textbf{$k$-Nearest Neighbors (KNN):} The number of nearest neighbors was varied among \{3, 5, 7, 9\}, using two different weighting schemes: \textit{uniform} (equal weight to all neighbors) and \textit{distance}-based (closer neighbors have greater influence). Additionally, two distance metrics were considered: \textit{Euclidean} and \textit{Manhattan}.
\end{itemize}

This set of algorithms was selected because they represent a diverse set of widely used ML paradigms, ranging from linear (SVM) and distance-based (KNN) approaches to tree-based and ensemble methods (DT, RF, LightGBM), allowing a comprehensive comparison of both simple and complex classifiers for this prediction task.

To ensure proper generalization and avoid overfitting the data, the dataset has been split into a training subset (80\%) and a test subset (20\%). All model training, as well as hyperparameter calculation, has been performed on the training set, while the final validation has been conducted using the test subset. Furthermore, this data division was carried out randomly but stratified, maintaining the same proportion of the classes to be predicted in both sets. These training and test sets have been the same for each of the experiments conducted to ensure that the comparison between different models and configurations is fair.

Cross-validation has also been performed on the training set in each of the experiments, obtaining average scores for different combinations of training and validation data. These scores have been used to decide which classifier is the best for our task and apply it to the test set. The different folds in the cross-validation were also constructed in a stratified manner.

Due to an imbalance in the ``Communication'' class (288 vs. 435 students in the ``Education'' class), the F1-score was adopted as the main evaluation metric instead of overall accuracy. The F1-score is the harmonic mean between precision (proportion of correctly predicted positive cases among all the predictions) and recall (proportion of actual positive cases that were correctly identified), providing a balanced measure of performance when class sizes are unequal. Subsequently, the performance of the final model will be calculated using accuracy. Additionally, considered models, detailed in the following section, have been applied with balanced class weights to address this problem simply and improve final generalization.

\subsection{Selection of the Most Relevant Features for Predicting the Knowledge Branch}

Regarding feature selection, different methods have been considered, and subsequently, they have been compared using this reduced subset of features with our best model. The methods considered are:

\begin{itemize}
    \item \textbf{SelectKBest (SKB)}: This method employs the ANOVA (f\_classif) scoring function to evaluate feature importance, selecting the top $k$ features based on their scores. In this case, the selected features were obtained using the dataset transformed by this method.
    \item \textbf{Forward Feature Selector (FFS)}: Using an RF model with specific hyperparameters obtained from the best model configuration for the training data, this method selects features sequentially in a forward manner and optimizes the selection based on the accuracy metric using stratified five-fold cross-validation.
    \item \textbf{Recursive Feature Elimination (RFE)}: Also based on the previously configured RF model, this method recursively eliminates features until the desired number of selected features is reached.
    \item \textbf{Feature Importance Based on DT}: Using a DT classifier, the importance of each feature was calculated, selecting the most relevant $n$ features according to the generated importance scores (corresponding to the first splits in the tree branches).
\end{itemize}

The feature selection techniques were chosen to combine complementary strategies: univariate statistical ranking (SKB), sequential wrapper-based optimization (FFS), recursive elimination (RFE), and model-based importance estimation (DT). This combination ensures that both individual feature relevance and inter-feature dependencies are considered in the conducted analysis.

Finally, in order to identify the optimal feature subset for predicting the field of knowledge, we will use as a baseline the best-performing model obtained from the experimentation aimed at optimizing this prediction.

\subsection{Regression of the Measured Competencies}

The final experiment will involve performing a regression to estimate the scores a student would achieve in each of the four assessed competencies: Knowledge, Skills, Attitudes, and Responsibility. As input variables, all the students' sociodemographic characteristics will be considered, along with their course, field of study, and the scores obtained in the remaining competencies. To conduct the experimentation, various regression models and different hyperparameter configurations have been considered. The algorithms and configurations are as follows:

\begin{itemize}
    \item \textbf{Linear Regression (LR)}: This model captures the linear relationship between predictors and the target variable, estimating how each input feature proportionally influences the predicted competency score. The model was evaluated by considering whether to include an intercept term in the equation. The inclusion of an intercept allows the model to better fit datasets where the dependent variable does not naturally pass through the origin.

    \item \textbf{Ridge Regression (RR)}: RR measures the same linear relationships as LR but adds a regularization term that penalizes large coefficients. Different values for the regularization strength parameter \(\alpha\) were explored, selecting from \{0.01, 0.1, 1.0, 10.0\}, where higher values impose stronger penalties on large coefficients to prevent overfitting. Additionally, different optimization solvers were tested:
    \begin{itemize}
        \item \textit{Singular Value Decomposition (SVD)}: A matrix factorization method that decomposes the design matrix into singular vectors and singular values.
        \item \textit{Least Squares (LSQR)}: An iterative method based on the conjugate gradient approach, solving the normal equations for least squares problems.
        \item \textit{SAGA}: A stochastic gradient-based method that updates coefficients in small batches, making it efficient for high-dimensional and sparse datasets.
    \end{itemize}
    
    \item \textbf{DT Regressor}: This model identifies non-linear, hierarchical relationships by recursively partitioning the feature space, creating threshold-based rules that predict competency scores. The tree-based model was evaluated using different maximum depths among \{\textit{unrestricted}, 5, 10, 20\}, where deeper trees can capture more complex patterns but may lead to overfitting. The minimum number of samples required to split a node was varied among \{2, 5, 10, 20\} to regulate tree complexity. Furthermore, two different criteria for measuring the quality of a split were considered:
    \begin{itemize}
        \item \textit{Squared Error}: Minimizes the variance within each node, leading to mean-squared-error minimization.
        \item \textit{Absolute Error}: Focuses on minimizing the mean absolute deviation, which is more robust to outliers.
    \end{itemize}
    
    \item \textbf{RF Regressor}: {The RF model extends the DT approach by combining multiple trees to capture more complex patterns and reduce variance}. This ensemble learning method was tested with different numbers of trees (estimators) among \{10, 50, 100\}, balancing computational cost and predictive performance. The features for each of the trees within the forest and the split criteria are the same as those considered for the DT Regressor.
\end{itemize}

These regression models were selected to capture different types of relationships between predictors and target variables: linear dependencies (LR, RR) and non-linear or hierarchical structures (DT, RF). Comparing them enables assessment of whether students’ competencies can be better explained by proportional or more complex interactions among variables.

Finally, the pipeline followed for the development of this experimental stage in terms of validation of the models remained identical to that of Section~\ref{sub:class}.

\section{Results}
\label{sec:results}
This section presents the main findings derived from the implementation of ML models to predict students’ MIL competencies. The results are organized into four subsections. First, we validate the questionnaire used for data collection and assess its reliability. Next, we examine the models’ classification performance across the two primary knowledge branches, Education and Communication. We then identify the most relevant features contributing to accurate classification and prediction. Finally, we analyze the performance of regression models used to estimate MIL competencies, highlighting key factors that influence their predictive accuracy.

\subsection{Preliminary Results: Validation and Reliability of the Questionnaire}

An analysis of the main psychometric properties of the instrument was carried out to verify its reliability and validity. The results showed an acceptable internal consistency (Cronbach's alpha of 0.77) and a high sampling adequacy (Kaiser-Meyer-Olkin = 0.85). Bartlett’s test of sphericity was significant ($\chi^2[190]=2995.42; p < .001$), confirming that the correlation matrix was suitable for factor analysis.
An exploratory factor analysis was then conducted using the principal component method with Varimax rotation. Although the eigenvalue-greater-than-one criterion initially suggested four factors, the fourth component was a residual and lacked conceptual coherence. Therefore, a three-factor solution was retained, explaining 66.86\% of the total variance. This configuration provided a meaningful and interpretable structure aligned with both empirical and theoretical considerations.
The interpretation of each factor was based on the content of the items with the highest loadings ($\geq 0.40$) and their conceptual relationship with existing frameworks of MIL and disinformation competence.
Table~\ref{tab:factor-analysis} shows the items that configure each factor.

\begin{table}[htbp]
\caption{Factor Analysis (Rotated Component Matrix)\label{tab:factor-analysis}}
\centering
\small
\begin{tabular}{cl|ccc}
\hline
\textbf{Id} & \textbf{Item} & \textbf{Factor 1} & \textbf{Factor 2} & \textbf{Factor 3} \\
\hline
K1 & I know what misinformation means. & 0.706 & & \\
K2 & I can identify the characteristics of & & & \\ & misinformation in a fake headline. & 0.757 & & \\
K3 & I differentiate the concept of misinformation & & & \\ & from similar terms such as "fake news",  & & & \\ & "rumors", and "false news". & 0.665 & & \\
K4 & I understand why misinformation spreads. & 0.694 & & \\
K5 & I can identify the risks of misinformation. & 0.679 & & \\ \hline
S1 & I differentiate true information from false & & &\\ & information based on the author or media outlet. & & 0.569 & \\
S2 & I verify the truthfulness of information with & & & \\ & different resources (news outlets, fact-checking & & & \\  & websites, social media, etc.). & & 0.472 & \\
S3 & I use various reliable sources of information & & &\\ & such as reputable newspapers, official websites, & & & \\ & and institutional pages to search for information & & & \\ & on topics that interest me. & & 0.430 & \\
S4 & I share information with my contacts without& & &\\ & verifying its truthfulness. & & 0.704 & \\
S5 & I report false information when I detect it on any & & &\\ & channel or platform. & & 0.316 & \\ \hline
A1 & I positively value the fight against misinformation & & &\\ & to guarantee freedom of expression. & & & 0.684\\
A2 & I am aware that misinformation goes against & & & \\ & the formation of a free and democratic citizenry. & & & 0.633 \\
A3 & I trust the information I receive through social media. & & & 0.628 \\
A4 & I believe that the media is trustworthy. & & & 0.786 \\
A5 & I give more credibility to information received from & & &\\ & a close contact (family, friends, etc.) than from & & & \\ & other sources. & & 0.595 & \\ \hline
R1 & I am aware that misinformation can affect the & & & \\ & exercise of my future professional practice. & & & 0.724\\
R2 & I perceive that the fight against misinformation& & & \\ & falls on professionals in my field. & & & 0.639\\
R3 & I recognize that misinformation is a necessary topic & & &\\ & in my education for the practice of my profession. & & & 0.644\\
R4 & I believe it is each person's responsibility to fight & & &\\ & against misinformation. & & & 0.545\\
R5 & I have the ethical and moral responsibility as a future& & &\\ & educommunicator to transmit truthful information. & & & 0.602 \\
\hline
\end{tabular}

\end{table}

Considering the content of the items that most saturated each factor, the following interpretation was reached:
\begin{itemize}
    \item \textbf{Factor 1: Knowledge about disinformation.} This dimension includes five items reflecting conceptual understanding of disinformation, i.e., its definition, characteristics, and differences from related terms such as fake news or hoaxes. It also encompasses awareness of the mechanisms that facilitate the spread of false content and the risks it poses to informed citizenship.
    \item \textbf{Factor 2: Skills and behaviors against disinformation.} Comprising six items, this factor captures the ability to verify information accuracy by consulting multiple and credible sources, such as established media outlets, fact-checking organizations, and official online platforms. It also reflects students’ discernment in evaluating information from social networks or close contacts.
    \item \textbf{Factor 3: Attitudes of commitment and responsibility towards disinformation.} This factor, composed of nine items, emphasizes ethical and civic engagement in countering disinformation. It involves recognizing the role of responsible communication in safeguarding freedom of expression and democratic participation, as well as acknowledging the professional duty, particularly among future educators and communicators, to disseminate truthful information in an ethical and reflective manner.
\end{itemize}

Together, these three empirically derived dimensions capture the multifaceted nature of students’ MIL competencies in the context of disinformation, integrating cognitive, behavioral, and attitudinal components in a coherent and statistically supported structure.

Once the construct structure was clarified, the internal consistency was calculated. The reliability coefficients in the three factors (Cronbach's alpha) reached a good threshold of .77, .79, and .86.

\subsection{Classification of the Knowledge Branch}\label{sec:res_class}

Table~\ref{tab:results_all_items} presents the classification performance of the evaluated models. The comparison is based on F1-score and accuracy for both training and test sets, providing insights into the models' generalization capabilities.

The results indicate that SVM achieved the highest performance, with a test F1-score of 0.815 and accuracy of 0.814, demonstrating both high predictive power and generalization ability. The RF Classifier and LightGBM followed closely, with F1-scores of 0.744 and 0.792, respectively. These ensemble models outperformed the DT Classifier, which had the lowest test F1-score (0.679) and accuracy (0.676), suggesting that a single-tree approach may be less effective for this classification task.

The KNN model showed moderate performance, with a test F1-score of 0.779 and accuracy of 0.779, surpassing the DT classifier but falling short of RF and SVM. Notably, KNN exhibited a low standard deviation in training accuracy (0.020), suggesting stable performance across different training splits.

Overall, SVM emerged as the most effective model, followed by ensemble methods (RF and LightGBM), while DT performed the worst. The results highlight the advantage of more complex models that leverage multiple decision boundaries over simpler tree-based or distance-based classifiers.

\begin{table}[htbp]
    \centering
    \small
    \caption{Results of the different models in predicting the knowledge branch, including the best configuration for each of the models. }
    \label{tab:results_all_items}
    \renewcommand{\arraystretch}{1.2}
    \begin{tabular}{ll|cccc|cc}
        \hline
        \multicolumn{2}{c}{\textbf{Model}} & \multicolumn{4}{c}{\textbf{Train (CV)}} & \multicolumn{2}{c}{\textbf{Test}} \\ 
        \hline
        &  & \textbf{Mean} & \textbf{Std} & \textbf{Mean } & \textbf{Std} & &  \\
        \textbf{Algorithm} & \textbf{Best Parameters} & \textbf{F1} & \textbf{F1} & \textbf{ Acc} & \textbf{Acc} & \textbf{F1} & \textbf{Acc} \\
        \hline
        & \textbf{\{'C': 1, 'gamma': 'scale',} & & & & & & \\ \textbf{SVM} & \textbf{'kernel': 'rbf'\}} &\textbf{ 0.734} & \textbf{0.032} & \textbf{0.782} & \textbf{0.033} & \textbf{0.815} & \textbf{0.814} \\ 
        \hline
        & \{'criterion': 'gini',& & & & & &\\ & 'max\_depth': 5, & & & & & &\\ DT Classifier & 'min\_samples\_split': 10\} & 0.662 & 0.041 & 0.704 & 0.040 & 0.679 & 0.676 \\ 
        \hline
        & \{'criterion': 'entropy',& & & & & &\\ & 'max\_depth': 5, & & & & & & \\ 
        & 'min\_samples\_split': 2, & & & & & & \\ RF Classifier & 'n\_estimators': 10\} & 0.705 & 0.061 & 0.773 & 0.037 & 0.744 & 0.745 \\
        \hline
        & \{'learning\_rate': 0.1, & & & & & & \\ & 'max\_depth': 5, & & & & & & \\  
        LightGBM & 'n\_estimators': 50\} & 0.683 & 0.058 & 0.765 & 0.043 & 0.792 & 0.793 \\ 
        \hline
        & \{'metric': 'euclidean', & & & & & & \\ & 'n\_neighbors': 5, & & & & & & \\  
        KNN & 'weights': 'uniform'\} & 0.658 & 0.034 & 0.747 & 0.020 & 0.779 & 0.779 \\ 
        \hline
    \end{tabular}
\end{table}

\subsection{Selection of the Most Relevant Features for Predicting the Knowledge Branch}

Feature selection plays a critical role in ML models, as it can significantly impact both model interpretability and predictive performance. This section presents the results of different feature selection methods applied to the dataset, aiming to identify the most relevant features for classification. Table \ref{tab:results_feature_selection} summarizes the selected features and the corresponding model performance metrics.

\begin{table}[htbp]
    \centering
    \small
    \caption{Results obtained for different feature selection methods}
    \label{tab:results_feature_selection}
    \renewcommand{\arraystretch}{1.2}
    \begin{tabular}{ll|cccc|cc}
        \hline
        \multicolumn{2}{c}{\textbf{Selector}} & \multicolumn{4}{c}{\textbf{Train (CV)}} & \multicolumn{2}{c}{\textbf{Test}} \\ 
        \hline
        \textbf{Method} & \textbf{Best Features} & \textbf{Mean F1} & \textbf{Std F1} & \textbf{Mean Acc} & \textbf{Std Acc} & \textbf{F1} & \textbf{Acc} \\ 
        \hline
        SKB & K1, K2, K3, & & & & & & \\ & Disinfo Training = False, & & & & & & \\ & Disinfo Training = True & 0.682 & 0.030 & 0.763 & 0.019 & 0.751 & 0.752 \\ 
        \hline
        \textbf{FFS} & \textbf{K3, A3, R5,} & & & & & & \\ & \textbf{Disinfo Training = False,} & & & & & & \\ & \textbf{Disinfo Training = True} & \textbf{0.665} & \textbf{0.039} & \textbf{0.760} & \textbf{0.021} & \textbf{0.756} & \textbf{0.759} \\ 
        \hline
        RFE & Academic Year, K1, K2, & & & & & & \\ & Disinfo Training = False, & & & & & & \\ & Disinfo Training = True & 0.691 & 0.027 & 0.756 & 0.017 & 0.752 & 0.752 \\ 
        \hline
        DT & Academic Year, K3, A4, S5, & & & & & & \\ & Disinfo Training = False & 0.705 & 0.057 & 0.761 & 0.045 & 0.738 & 0.738 \\ 
        \hline
    \end{tabular}
\end{table}

The feature selectors exhibit notable variations in the features they prioritize. The SKB method selected three of the Knowledge category items, K1, K2, and K3, along with the Disinfo Training condition in both its False and True states. This suggests that these specific features hold significant discriminatory power according to univariate statistical tests.
The FFS identified a different subset, incorporating K3, A3, and R5, while also retaining the Disinfo Training condition in both states. Notably, the inclusion of A3 and R5 suggests that this method captures additional dependencies between features that may not be evident through individual statistical scoring.
The RFE approach selected a broader range of features, including Academic Year, K1, and K2, along with the Disinfo Training condition. The presence of academic year indicates that RFE might emphasize structural data attributes that influence model performance.
Finally, the DT method produced a unique feature subset, selecting Academic Year, K3, A4, and S5, while disregarding the Disinfo Training = True feature. This difference implies that the DT structure favors hierarchical relationships among features rather than purely statistical significance.

\begin{figure}
    \centering
    \includegraphics[width=\columnwidth]{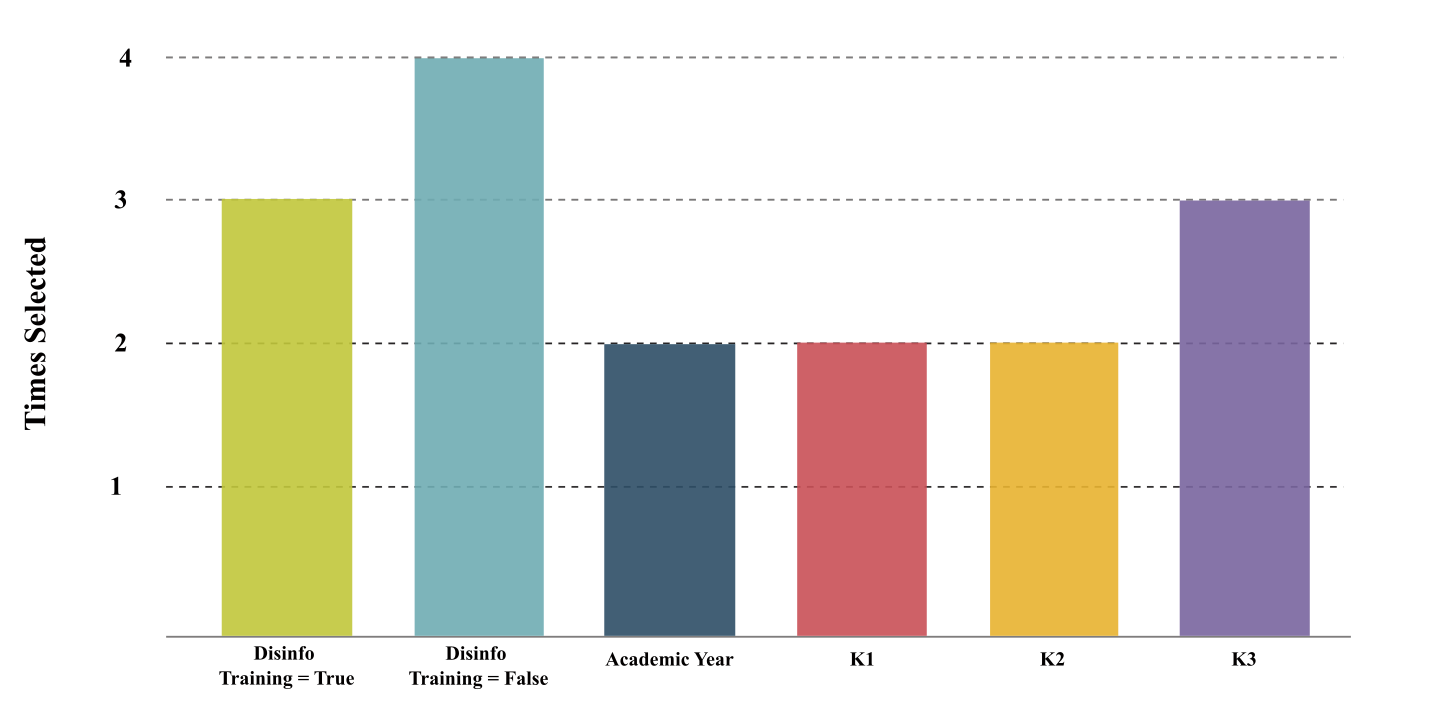}
    \caption{Most useful features selected by the different feature selection methods in order to predict the knowledge branch}
    \label{fig:top_features}
\end{figure}

Figure \ref{fig:top_features} further highlights the importance of specific features across different selection methods. Notably, Disinfo Training = False was the most frequently selected feature, appearing in all four selection methods, followed closely by Disinfo Training = True, which was chosen three times. This suggests that the disinformation training condition is a key determinant in the classification process. Additionally, K3 was selected three times, reinforcing its importance, whereas Academic Year, K1, and K2 were each chosen twice, indicating their moderate but relevant contribution to predictive performance.

Among all feature selection methods, FFS achieved the highest test F1-score (0.756) and accuracy (0.759), indicating that its selected feature set contributed most effectively to model generalization. The DT-based selection followed closely with an F1-score and accuracy of 0.738, suggesting that its feature set was also effective but slightly less optimal for this classification task.
Interestingly, SKB and RFE exhibited similar performance, with both reaching an F1-score of 0.751–0.752 and test accuracy of 0.752, reinforcing the idea that simpler selection methods based on statistical significance and elimination heuristics can still yield competitive results.

Overall, these findings indicate that FFS provided the most robust feature selection strategy, likely due to its ability to iteratively assess combinations of features, leading to improved generalization. This highlights the importance of considering feature interactions rather than relying solely on individual statistical measures when optimizing feature selection for classification tasks. The frequency analysis in Figure \ref{fig:top_features} further supports this conclusion, as FFS incorporated highly recurrent features such as K3 and Disinfo Training variables, which were also selected by other methods.

\subsection{Regression of the Measured Competencies}
Table~\ref{tab:regression_results} presents the predictive performance of the regression models for the different objective variables. The results are evaluated based on the Root Mean Squared Error (RMSE) for both the training phase, using cross-validation, and the test phase. Lower RMSE values indicate better predictive accuracy.

\begin{table}[htbp]
    \centering
    \small
    \caption{Regression performance metrics for the evaluated models.}
    \label{tab:regression_results}
    \renewcommand{\arraystretch}{1.2}
    \begin{tabular}{ll|cc|c}
        \hline
        \multicolumn{2}{l}{\textbf{Regression Target and Selected Model}} & \multicolumn{2}{c}{\textbf{Train (CV)}} & \textbf{Test} \\ 
        \hline
        \textbf{Objective Variable} & \textbf{Best Model} & \textbf{Mean RMSE} & \textbf{Std MSE} & \textbf{Test RMSE} \\ 
        \hline
        Knowledge & LR \{'fit\_intercept': True\} & 2.201 & 0.311 & 2.287 \\ 
        \hline
        & RF \{'criterion': 'squared\_error', & & & \\ & 'max\_depth': 5, & & & \\  
        & 'min\_samples\_split': 20, & & & \\ Skills & 'n\_estimators': 100\} & 1.759 & 0.455 & 1.835 \\ 
        \hline
        Attitudes & RR \{'alpha': 10.0\} & 1.686 & 0.354 & 1.598 \\ 
        \hline
        & RF \{'criterion': 'squared\_error', & & & \\ & 'max\_depth': 5, & & & \\ & 'min\_samples\_split': 2, & & & \\ Responsibility & 'n\_estimators': 100\} & 2.088 & 0.715 & 2.064 \\  
        \hline
    \end{tabular}
\end{table}

The results indicate that the best model varies depending on the objective variable. Knowledge was best predicted by LR, with a test RMSE of 2.287. This suggests that a simple linear relationship between the input features and the target variable is sufficient for accurate predictions. Skills and Responsibility were best modeled by RF, achieving test RMSE values of 1.835 and 2.064, respectively. The ensemble approach of RF likely allowed for better handling of complex, nonlinear relationships in these cases. Attitudes were best captured by RR, with a test RMSE of 1.598. The use of regularization in RR likely helped in controlling model complexity and improving generalization.

Among all models, RR for Attitudes achieved the lowest test RMSE (1.598), indicating the best predictive performance overall. The RF models exhibited higher variability in their training RMSE, especially for Responsibility, with a standard deviation of 0.715, which suggests sensitivity to different training subsets. In contrast, LR and RR showed more stable performance across cross-validation folds.

\section{Discussion and Limitations}\label{sec:discussion}

The preceding section presented the main findings of the machine learning analyses. This section discusses their implications, situates them within the existing literature, and addresses the study’s limitations.

From an educational perspective, the findings presented in Section~\ref{sec:res_class} highlight the effectiveness of ML tools in assessing MIL competencies. The SVM model performed the best, suggesting that more complex approaches are essential for accurately evaluating MIL skills, as they can consider multiple variables and interactions. The strong performance of RF and LightGBM models further emphasizes the importance of using methods that can handle complex relationships in the data. These models are particularly useful in assessing students' knowledge, skills, and attitudes toward misinformation, which is increasingly vital in modern education. On the other hand, the lower performance of the DT model highlights the limitations of simpler approaches, indicating that more robust models are needed for assessing MIL in future educators and communicators. Finally, while KNN had lower accuracy, its stability suggests that it could still be valuable in smaller groups or less variable contexts.

Comparing these outcomes with prior works reveals several points of convergence. Studies such as Reddy et al.~\citep{Reddy19} and Mohd Nadzir and Abu Bakar~\citep{MohdNadzir23} also reported the superiority of non-linear classifiers, particularly SVM, in educational prediction tasks. Similarly, our results support the notion that ensemble models outperform single-tree approaches when modeling students’ perceptual data, due to their robustness to noise and ability to generalize across heterogeneous populations.

The analysis of feature selection also provides meaningful insights. The feature selection methods demonstrated that variables related to disinformation training were consistently important across all methods, underlining the critical role that prior training in misinformation plays in shaping students' understanding and responses. The results also show that features such as knowledge items (K1, K2, K3) and attitudes toward misinformation (A3, R5) were influential in accurately predicting knowledge outcomes. Additionaly, the highest performance of the FFS method suggests that educational approaches should not only focus on standalone knowledge but also incorporate a more holistic understanding of how various factors, like academic year or previous training, interact to influence students' competencies. The variability in feature importance across methods further suggests that educational systems should take a multifaceted approach when assessing and addressing MIL, ensuring that all relevant factors are considered to improve training and education on misinformation.

Finally, regarding the regression of the competencies, the results emphasize the importance of choosing appropriate regression models for different competencies in education. LR is effective for predicting knowledge, as it handles simple relationships well. For skills and responsibility, RF excels due to its ability to manage complex, nonlinear relationships. RR performs best for predicting attitudes, thanks to its stability through regularization. These findings suggest that selecting the right model based on the nature of the competency can enhance the accuracy of educational assessments.

Despite these contributions, some limitations must be acknowledged. First, the study relies on self-reported data, which may be subject to social desirability bias or inaccuracies in participants’ self-perception. Second, ML models require large volumes of data to generate accurate predictions, which can be challenging in studies with relatively small or heterogeneous samples. Moreover, the sample was restricted to students in education and communication degrees from a single geographic region, which may constrain the generalizability of the results. Future research could address these issues by combining self-perception data with performance-based assessments and expanding the sample to include other disciplines and institutions.

\section{Conclusions}
\label{sec:conclusions}

This study demonstrates the potential of ML tools in assessing MIL competencies among students. Key findings show that more complex models, such as SVM, RF, and LightGBM, provide better insights into students’ abilities, highlighting the need for a holistic approach that considers multiple factors, such as academic year and prior training, when evaluating MIL skills.

The research emphasizes that educational systems should integrate various elements of students' backgrounds to improve their understanding of misinformation. It suggests that incorporating ML tools and considering a variety of student-related factors can significantly improve how misinformation is addressed within educational frameworks, ultimately fostering more critical and informed students.

Importantly, this study contributes to the international academic debate on MIL in the context of disinformation by bridging theoretical understanding and advanced methodological approaches. While prior research has largely relied on survey-based assessments or traditional statistical analyses, this study demonstrates how ML can be used to model and predict MIL competencies, providing a more nuanced and data-driven understanding of the factors influencing students’ skills. Furthermore, by examining sociodemographic variables such as gender, academic year, and prior training, the study highlights patterns and disparities that have been inconsistently addressed in the existing literature, thereby offering new insights into how these factors shape MIL development across diverse educational contexts.

Finally, the findings provide practical implications for designing targeted interventions and evidence-based educational strategies. Unlike previous studies that have primarily focused on general MIL competencies, this work explicitly addresses MIL against disinformation, demonstrating the relevance of predictive analytics for tailoring educational initiatives. By linking ML-based predictions with theoretically grounded MIL dimensions, the study extends the current literature on MIL, informing both policy recommendations and pedagogical practices aimed at enhancing students’ ability to critically navigate and respond to disinformation in digital environments.

Future research should investigate the generalizability of these findings across diverse regions, educational programs, and cultural contexts. Moreover, the integration of more advanced ML techniques, such as deep learning, has the potential to further improve prediction accuracy. Subsequent studies could also examine the effectiveness of targeted interventions informed by these predictive models in enhancing students’ MIL competencies against disinformation.

\vspace{6pt} 





\authorcontributions{{Conceptualization},  J.M.A.-L., M.B.-F, C.E.G.-R., A.Z. and E.Y.-B.; Methodology, J.M.A.-L., M.B.-F, C.E.G.-R. and E.Y.-B.; Resources, J.M.A.-L., M.B.-F and E.Y.-B.; Data curation, J.M.A.-L., M.B.-F and E.Y.-B.; Writing---original draft preparation, J.M.A.-L., M.B.-F, C.E.G.-R. and E.Y.-B.; Writing---review and editing, J.M.A.-L., M.B.-F, C.E.G.-R., A.Z. and E.Y.-B.; Supervision, M.B.-F and E.Y.-B. All authors have read and agreed to the published version of the manuscript.}

\funding{This work was conducted with the support of the project ''Media and Digital Literacy in Young People and Adolescents: Diagnosis and Educational Innovation Strategies to
Prevent Risks and Promote Good Practices on the Internet,´´
funded by the Department of Universities of the Government
of Cantabria (Spain). Additionally, these results are part of
UNITE: University Network for Inclusive and digiTal Education, project funded by the European Union within the
Erasmus+ programme (call 2023), Key Action 2: KA220
Cooperation Partnerships for higher education (2023-1-IT02-
KA220-HED-0001621181).}

\institutionalreview{
The research adhered strictly to ethical
regulations, including Spain’s Organic Law 3/2018 on the Protection of Personal Data and the principles outlined in the
Declaration of Helsinki. Ethical approval was obtained from
the university’s institutional ethics committee (CEIH-25-54). 
}

\informedconsent{Informed consent was obtained from all subjects involved in the study.}

\dataavailability{The raw data supporting the conclusions of this article will be made available by the authors on request.}

\acknowledgments{José Manuel Alcalde-Llergo enrolled in the National PhD in Artificial Intelligence, XXXVIII cycle, course on Health and Life Sciences, organized by Università Campus Bio-Medico di Roma. He is also pursuing his doctorate with co-supervision at the Universidad de Córdoba (Spain), enrolled in its PhD program in Computation, Energy and Plasmas.}

\conflictsofinterest{The authors declare no conflicts of interest.} 



\begin{adjustwidth}{-\extralength}{0cm}

\reftitle{References}


\bibliography{references}

%


\PublishersNote{}
\end{adjustwidth}
\end{document}